\documentclass[a4paper,11pt]{article}
\usepackage{pos}
\usepackage{amsmath}
\usepackage{dsfont}
\usepackage{bbm}
\usepackage{cleveref}
\usepackage{subcaption}
\usepackage[utf8]{inputenc}
\usepackage{mathrsfs}
\usepackage{amsfonts}
\usepackage{wrapfig}

\let\OLDthebibliography\thebibliography
\renewcommand\thebibliography[1]{
  \OLDthebibliography{#1}
  \setlength{\parskip}{0pt}
  \setlength{\itemsep}{0pt plus 0.3ex}
}

\DeclareMathOperator{\tr}{tr}

\newcommand{\fm}{\text{fm}}
\newcommand{\dif}{\mathrm{d}}

\newcommand{\flow}{\mathrm{fl}}

\newcommand{\chihat}{\hat\chi}
\newcommand{\Ohat}{\hat{O}}
\newcommand{\rhat}{\hat r}
\newcommand{\ahat}{\hat a}
\newcommand{\tflow}{t_{\flow}}

\title{Gauge field smearing and controlled continuum extrapolations}
\ShortTitle{Gauge field smearing and controlled continuum extrapolations}

\author*[a,b]{Andreas Risch}

\affiliation[a]{Department of Physics, University of Wuppertal, Gaussstr. 20, 42119 Wuppertal, Germany}
\affiliation[b]{John von Neumann-Institut f{\"u}r Computing NIC, Deutsches Elektronen-Synchrotron DESY,\\
Platanenallee 6, 15738 Zeuthen, Germany}

\emailAdd{andreas.risch@uni-wuppertal.de}

\abstract{Two popular methods to reduce discretisation effects are Symanzik improvement and gauge field smearing in the Dirac operator. Tree-level $O(a^2)$-improved Wilson fermions can be obtained from $O(a)$-improved Wilson fermions by adding one dimension-6 operator to the action. For gauge field smearing one wants to avoid the situation when too much smearing leads to uncontrolled continuum extrapolations as the short distance behaviour is mutilated. We focus on the gradient flow formalism as it allows to study both smearing and physical flow. We investigate the effect of smearing on the scaling towards the continuum limit in pure gauge theory on the example of Creutz ratios, which provide a measure of the physical forces felt by the fermions. For suitable smearing strengths we also investigate the change when the Wilson gradient flow is replaced by stout smearing.}

\FullConference{The 40th International Symposium on Lattice Field Theory (Lattice 2023)\\
July 31st - August 4th, 2023\\
Fermi National Accelerator Laboratory\\}

\begin{document}
\maketitle

\section{Introduction}

A reduction of lattice artefacts is beneficial for a more reliable continuum extrapolation, in particular of short distance observables. Two popular methods to alter discretisation effects are Symanzik improvement and UV filtering. Assuming an (on-shell) $O(a)$-improved Wilson fermion action as e.g. used by the CLS effort~\cite{Bruno:2014jqa}, a tree-level $O(a^2)$-improvement can be achieved by considering the lattice discretisation of one additional irrelevant dim-6 operator~\cite{Sheikholeslami:1985ij} in the Lagrange density, namely $\delta\mathscr{L}=-\frac{a^{2}}{6}\sum_{\mu}\overline{\Psi}\gamma_{\mu}\nabla_{\mu}\Delta_{\mu}\Psi$, which breaks Lorentz symmetry. The corresponding lattice actions belong to the class of W2/D234 fermion actions~\cite{Alford:1996nx}. Considering an $O(a)$-improved lattice action, the leading order lattice artefacts in the SymEFT expansion~\cite{Husung:2019ytz,Husung:2021mfl} can be written as $\mathcal{O}(a)-\mathcal{O}(0)\sim\sum_{i}\hat{c}_{i}\,M_{i}\,a^2 (\alpha_{\mathrm{s}}(a^{-1}))^{\Gamma_{i}}$, where $M_{i}$ and $\Gamma_{i}$ are observable dependent. $\alpha_{\mathrm{s}}(a^{-1})\sim(-\log(a\Lambda))^{-1}$ gives rise to logarithmic corrections in the lattice spacing, where $\Lambda$ is the intrinsic scale of the theory. When modifying the Lagrange density as described, the $\Gamma_{i}$ are shifted by one unit, i.e. $\Gamma_{i} \rightarrow \Gamma_{i}+1$, which leads to a parametric reduction of the lattice artefacts by one power of $\alpha_{\mathrm{s}}(a^{-1})\sim(-\log(a\Lambda))^{-1}$~\cite{Husung:2021mfl}.

UV-filtering is based on the application of  four-dimensional smearing on the gauge fields. Several smearing algorithms have been developed, e.g. HYP~\cite{Hasenfratz:2001hp}, Stout~\cite{Morningstar:2003gk}, HEX~\cite{Capitani:2006ni} and gradient flow~\cite{Narayanan:2006rf,Luscher:2010iy} smearing. Considering a smearing transformation $\mathcal{S}:U\mapsto \mathcal{S}[U]$, the Dirac operator is evaluated on smeared gauge fields, i.e. the action is altered into $S[U,\Psi,\overline{\Psi}] = S_{\mathrm{g}}[U] + \overline{\Psi}\,D[\mathcal{S}[U]]\,\Psi$. This procedure yields several advantages: The likelihood of finding small eigenvaluesof $D$, which is related to exceptional configurations, is reduced. In~\cite{Hasenfratz:2007rf,Durr:2008rw} the Wilson Dirac operator defined with nHYP/stout smeared gauge links could be shown to exhibit a spectrum with a well-defined spectral gap. This is particularly helpful for the simulation of mass non-degenerate quarks as the fermion determinant is not necessarily positive in such a scenario~\cite{Mohler:2020txx}. Gauge field smearing also alters improvement coefficients and renormalisation constants. For $c_{\mathrm{SW}}$ it was observed that the latter approaches its tree-level value when gauge field smearing is applied~\cite{Hasenfratz:2007rf}. The amount of renormalisation of the local vector current is also reduced. However, the application of too much smearing may significantly alter the UV structure of the lattice theory and therefore continuum extrapolations based on data from insufficiently small lattice spacings may become unreliable. It is therefore relevant to study what smearing strengths still allow for controlled continuum extrapolations. As a first step towards a smeared action setup including fermions we will study smeared observables $\langle O_{\mathcal{S}}[U] \rangle = \langle O[\mathcal{S}[U]] \rangle$ in pure gauge theory. We will investigate the influence of smearing on continuum extrapolations of Creutz ratios~\cite{Creutz:1980wj}, which provide a measure of the physical forces felt by the fermions caused by the gauge field. In a previous account of this effort~\cite{Risch:2022vof} the computational setup is described in more detail. Creutz ratios in combination with APE smearing and the Wilson flow were also studied in a determination of the string tension in~\cite{Okawa:2014kgi}.

\section{The gradient flow formalism, gradient flow smearing and physical gradient flow}
\label{sec:gradientflow}

In this work, we focus on the gradient flow formalism~\cite{Luscher:2010iy} as a smearing procedure. Details on the applied lattice discretisation and on the integration of the gradient flow equation can be found in~\cite{Risch:2022vof}. In the following, we consider two scenarios in which we will apply the gradient flow to the gauge field. In the first scenario, which we refer to as gradient flow smearing, the gradient flow time $\tflow$ and hence the smearing radius vanish in the continuum limit such that the continuum theory is unaltered. This is achieved by fixing the gradient flow time in lattice units, i.e. $\frac{8\tflow}{a^{2}} = \mathrm{const}$. The second scenario, in which the flow time is fixed in physical units, i.e. $\tflow / t_0= \mathrm{const}$, we refer to as a physical gradient flow. $t_0$ may be any physical scale of the theory, in particular the reference flow time defined in~\cite{Luscher:2010iy}. In this scenario the continuum theory is altered. This alteration of the observable's continuum limit can also be understood as a modification of the observable's definition.

\section{Combined continuum extrapolation and small flow time expansion}
\label{sec:combinedextrapolation}

In the following, we consider a dimensionless observable $\Ohat$ that does not require a renormalisation and hence is finite in the continuum limit. We will understand this observable as a function of the dimensionless lattice spacing parameter $\ahat\equiv\frac{a}{\sqrt{8t_{0}}}$ and the flow time parameter $\varepsilon=\frac{\tflow}{t_{0}}$. The finiteness of the observable entails the interchangeability of the continuum limit and the zero flow time limit, i.e. $\lim_{\ahat \rightarrow 0}\lim_{\varepsilon \rightarrow 0} \Ohat = \lim_{\varepsilon \rightarrow 0}\lim_{\ahat \rightarrow 0} \Ohat$. For such finite observables the two scenarios discussed in \cref{sec:gradientflow} have a common limit where both $a=0$ and $t_\flow=0$. Therefore, a combined Symanzik expansion in the lattice spacing and a small flow time expansion is possible and well-defined. As our investigation has an intermediate precision, we neglect logarithmic effects both in the lattice spacing~\cite{Husung:2021mfl} and in the flow time~\cite{Luscher:2010iy}. The double expansion of the observable then reads
\begin{align}
\Ohat &= \sum_{i,j \geq 0}c_{ij} \ahat^{i}\varepsilon^{j}.\label{eq:obsdoubleexpansionflow}
\end{align}
Evaluating this expression at $\ahat=0$, it becomes obvious that the observable's continuum limit can be altered by a physical gradient flow as $\Ohat = c_{00} + \sum_{j>0}^{n}c_{0j} \varepsilon^{j}$, where $c_{00}$ denotes the continuum limit at vanishing flow time. In this work, we are primarily interested in the effect of smearing on the continuum extrapolation. In order to demonstrate that this expansion can also be used to describe the observable's lattice spacing dependence at fixed smearing strength $\frac{8\tflow}{a^{2}}$, we observe that the latter is parametrised by $\frac{\varepsilon}{\ahat^{2}}=\frac{8\tflow}{a^{2}}$. We can therefore rewrite the expansion as a function of the lattice spacing and the smearing strength:
\begin{align}
\Ohat &= \sum_{i,j \geq 0}c_{ij} \ahat^{i+2j}\Big(\frac{\varepsilon}{\ahat^{2}}\Big)^{j} = \sum_{i,j \geq 0}c_{ij} \ahat^{i+2j}\Big(\frac{8\tflow}{a^{2}}\Big)^{j}. \label{eq:obsdoubleexpansionsmearing}
\end{align}
Evaluating the smearing expansion at $\ahat=0$ yields $\Ohat = c_{00}$, i.e. the continuum limit is independent of the smearing strength by construction. The main advantage of this double expansion is that data from various small $\ahat\equiv\frac{a}{\sqrt{8t_{0}}}$ and $\varepsilon=\frac{\tflow}{t_{0}}$ can be combined to determine the coefficients $c_{ij}$, which then allow to reconstruct the lattice spacing dependence for any (sufficiently small) $\ahat$ and $\varepsilon$.

\section{Creutz ratios and gradient flow}

For a study in pure gauge theory Creutz ratios~\cite{Creutz:1980wj} are suitable observables as they possess a finite continuum limit. Creutz ratios are constructed from planar rectangular Wilson loops $W(r, t)\!\equiv\!\langle \tr( P \exp(\oint_{\gamma(r,t)} \dif x_{\mu}A_{\mu}(x))) \rangle$, which are obtained from the gauge field by a path-ordered integral along a rectangular closed path $\gamma(r,t)$. On the lattice these objects are discretised as
\begin{gather}
W(r,t) = \Big\langle \tr\Big(\prod_{(x,\mu)\in\gamma(r,t)} U_{\mu}(x)\Big) \Big\rangle.
\end{gather}
In the continuum, Creutz ratios are obtained from Wilson loops by $
\chi(r, t) \equiv -\frac{\partial}{\partial t}\frac{\partial}{\partial r} \ln(W(r, t))$. To achieve $O(a^{2})$ lattice artefacts we discretise the latter making use of central differences~\cite{Okawa:2014kgi}:
\begin{gather}
\chi\Big(t+\frac{a}{2}, r+\frac{a}{2}\Big) \equiv \frac{1}{a^{2}}\ln\Big(\frac{W(t+a, r)\cdot W(t, r+a)}{W(t, r) \cdot W(t+a, r+a)}\Big). \label{eq:creutzlat}
\end{gather}
From these quantities the static quark anti-quark force can be extracted in the limit of an infinite time extent, $\chi(r,t) \rightarrow F_{\mathrm{\overline{q}q}}(r)$ for $t\rightarrow\infty$. In the following discussion, we will only focus on diagonal Creutz ratios $\chi(r, t)$ with $r=t$, which we abbreviate as $\chi(r)\equiv\chi(r, r)$.

Our investigation is based on $\mathrm{SU}(3)$ Yang Mills theory gauge ensembles~\cite{Husung:2017qjz} using the Wilson plaquette action. Temporal open boundary conditions~\cite{Luscher:2011kk} are imposed to alleviate topology freezing. The reference flow time $t_{0}$~\cite{Luscher:2010iy} is used as a scale to construct dimensionless quantities. The lattice spacing varies between $0.08\,\fm$ and $0.02\,\fm$ and the spatial extent between $1.9\,\fm$ and $2\,\fm$. More details on the lattice setup can be found in~\cite{Risch:2022vof}.  We compute diagonal Creutz ratios in lattice units $(\chi\cdot a^{2})(\frac{r}{a})$ for various half integer distances $\frac{r}{a}=1.5,2.5,\ldots$ on gauge configurations which the gradient flow was applied to. We use $t_0$ to define dimensionless Creutz ratios, i.e. we analyse $\chihat \equiv \chi\cdot 8 t_{0}$ as a function of $\rhat \equiv\frac{r}{\sqrt{8t_{0}}}$. The two scenarios for scaling $t_\flow$ are implemented as
\begin{equation}
\label{eq:scenarios}	
\frac{8t_\flow}{a^2} =
\begin{cases}
0,\; 0.25,\; 0.5,\; \ldots,\; 2, \; 2.5, \ldots, \;3.5,\;4,\; 5,\; 6,\; 7, \; 8 & \text{smearing}\\
\frac{8t_0}{a^2} \times0.0146 \times j\,,\;\; j \in\{0,\;1,\;\ldots ,\;4\} & \text{physical flow}.
\end{cases}
\end{equation}
The computation is based on the openQCD~\cite{Luscher:openQCD} package and utilises B. Leder's program for measuring Wilson loops~\cite{Leder:wloop,Donnellan:2010mx}. For the data analysis the python3 package pyobs~\cite{Bruno:pyobs} is used, which implements the $\Gamma$-method~\cite{Wolff:2003sm} for Markov Chain Monte Carlo error estimation.

\begin{figure}
\centering
\includegraphics[width=0.495\textwidth]{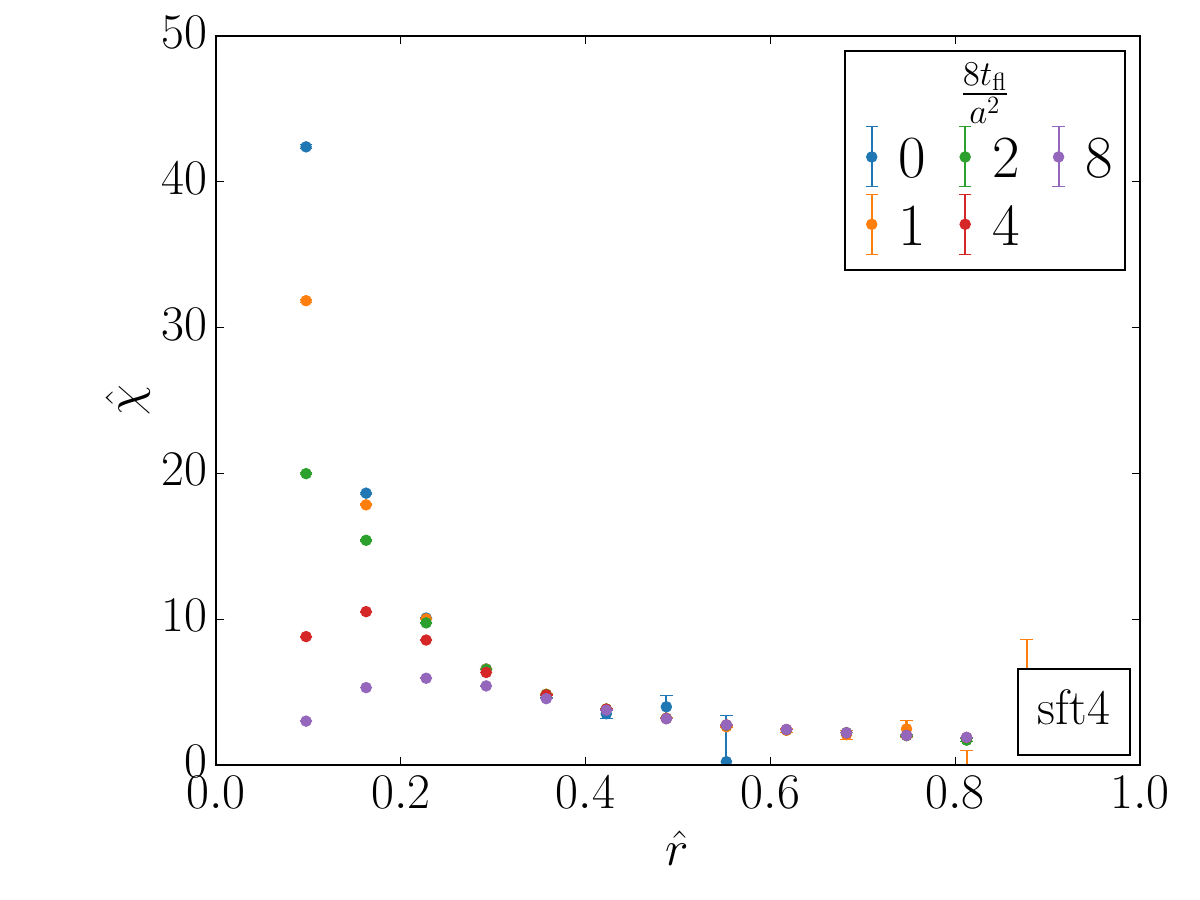}
\includegraphics[width=0.495\textwidth]{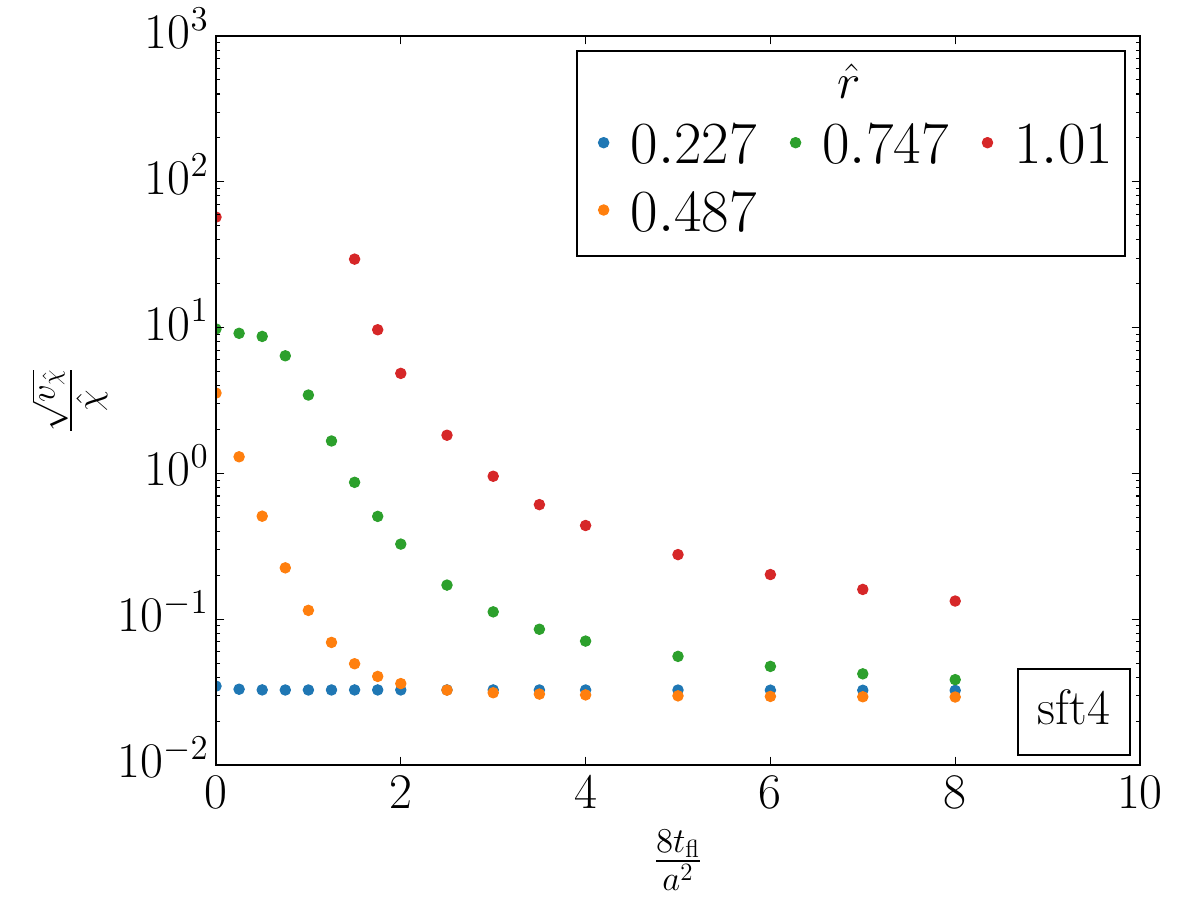}
\caption{Dimensionless Creutz ratio $\hat\chi$ and relative variance $\frac{\sqrt{v_{\hat\chi}}}{\hat\chi}$ as functions of the flow time $\frac{8t_{\flow}}{a^{2}}$ and the distance $\hat r$ on sft4.\label{fig:smearing}}
\end{figure}

Smearing is commonly applied to reduce UV fluctuations in gauge fields. The latter reduction also influences the variance of observables. In \cref{fig:smearing} the dimensionless diagonal Creutz ratio $\chihat$ and its relative variance $\frac{\sqrt{v_{\hat\chi}}}{\hat\chi}$ are displayed as functions of the distance $\rhat$ and the smearing strength $\frac{8t_{\flow}}{a^{2}}$ evaluated on the ensemble sft4. We observe that the $\sim \frac{1}{r^2}$ short distance behaviour is smoothed by the gradient flow at distances $r \lessapprox \sqrt{8t_\flow}$. Already from this qualitative picture we can conclude that in the smearing scenario the path to continuum and hence lattice artefacts are altered. This effect becomes smaller at larger distances where the smearing has less impact. Further, we observe that the relative variance of the Creutz ratio $\frac{\sqrt{v_{\hat\chi}}}{\hat\chi}$ grows with growing distances, i.e. there is a noise problem at large distances. Applying gradient flow smearing, the relative variance shrinks with growing flow time at all distances~\cite{Okawa:2014kgi}. However, it only shrinks to a certain level which seems to be almost independent of the distance $\rhat$ at which the Creutz ratio is evaluated, i.e. smearing the gauge fields does not lead to an arbitrarily large reduction of the relative variance.

\begin{figure}
\centering
\begin{minipage}[b]{0.49\textwidth}
\centering
\includegraphics[width=\textwidth]{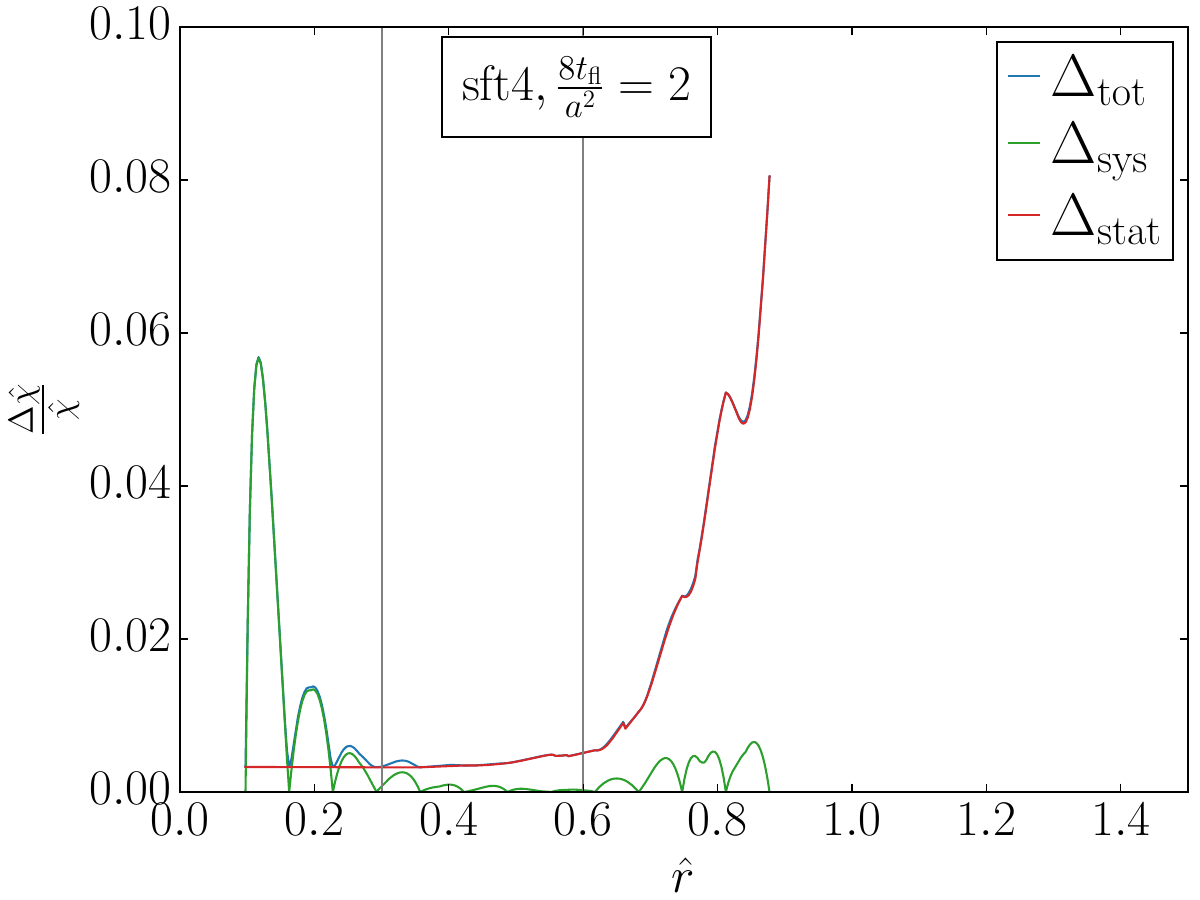}
\caption{Relative variance $\frac{\sqrt{v_{\hat\chi}}}{\hat\chi}$ as a function of the distance $\rhat$ on the ensemble sft4 with a gradient flow time $\frac{8 \tflow}{a^{2}}=2$.}
\label{fig:interpolation}
\end{minipage}
\begin{minipage}[b]{0.02\textwidth}
\end{minipage}
\begin{minipage}[b]{0.49\textwidth}
\centering
\includegraphics[width=\textwidth]{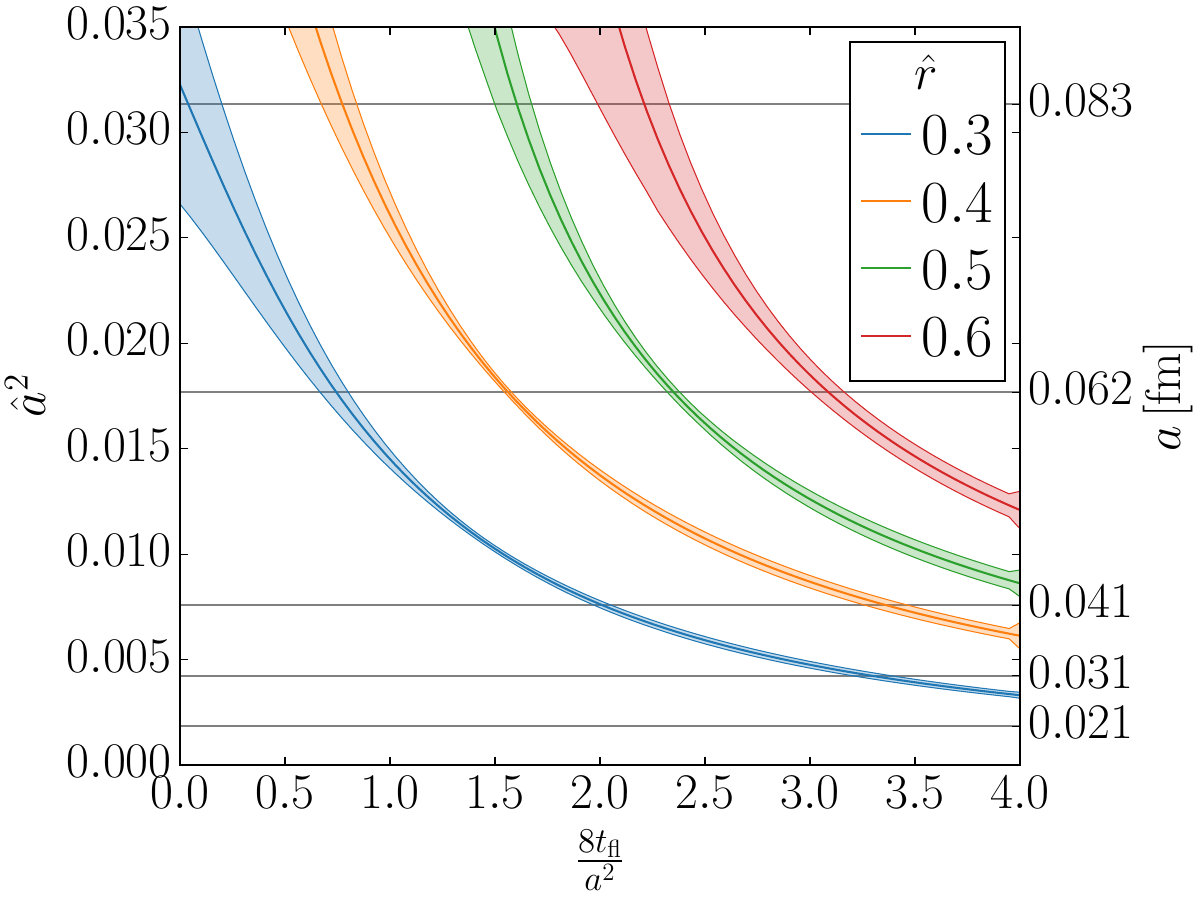}
\caption{Location of the maximum of $\chihat(\ahat)$ as a function of the smearing strength $\frac{8\tflow}{a^{2}}$ for several distances $\rhat\equiv\frac{r}{\sqrt{8t_{0}}}$.}
\label{fig:peaklocation}
\end{minipage}
\end{figure}

To be able to perform a continuum extrapolation of a Creutz ratio $\chihat(\rhat)$ evaluated at a fixed distance $\rhat$, $\chihat(\rhat)$ has to be known on all ensembles at $\rhat$. Therefore, we have to interpolate $\chi\cdot a^{2}$ as a function of $\frac{r}{a}$. Details on the used interpolation models are described in~\cite{Risch:2022vof}. We use the difference of different interpolations as a systematice error $\Delta_\mathrm{sys}$. In the following, we focus on the region $0.3 \leq\rhat\leq 0.6$ ($0.14\,\fm \leq r \leq 0.28\,\fm$), where lattice artefacts are not uncontrollably large and the statistical as well as the systematic uncertainty related to the interpolation are sufficiently small, c.f. \cref{fig:interpolation}.

\section{Influence of the smearing strength on the continuum extrapolation}

At fixed distance $\rhat$ we perform a global continuum extrapolation as discussed in \cref{sec:combinedextrapolation} combining data from both scenarios, smearing and physical gradient flow, c.f. \cref{eq:scenarios}. We construct a fit ansatz for the extrapolation by truncating the expansion \cref{eq:obsdoubleexpansionsmearing}, obtaining
\begin{align}
\chihat_{\mathrm{tr}} &= c_{00} + c_{20}\,\Big(1+\frac{c_{01}}{c_{20}}\frac{8\tflow}{a^{2}}\Big)\,\ahat^{2} + c_{40}\,\Big(1 + \frac{c_{21}}{c_{40}}\frac{8\tflow}{a^{2}} + \frac{c_{02}}{c_{40}}\Big(\frac{8\tflow}{a^{4}}\Big)^2\Big)\,\ahat^{4}. \label{eq:fitsmearing}
\end{align}
Further details on the extrapolation are described in~\cite{Risch:2022vof}. As a loose criterion for a controlled continuum extrapolation we may demand for a monotonous extrapolation, which we will apply in the following to the smearing scenario. That criterion will limit the smearing strength and the lattice spacing of the coarsest ensemble considered. To study monotony we track the location $\ahat_\mathrm{peak}^{2}$ of the extremum of $\chihat$ as a function of $\frac{8t_{\flow}}{a^{2}}$ at various distances $\rhat$. Then, all lattice spacings $\ahat^{2}<\ahat_\mathrm{peak}^{2}$ belong to a monotonous extrapolation. In \cref{fig:peaklocation} the location of the peak $\ahat_\mathrm{peak}^2$ is plotted as a function of the smearing radius for various distances $\rhat$. Given a minimum distance $\rhat$ for which the continuum extrapolation is monotonous, all points $\big(\frac{8\tflow}{a^{2}},\ahat^{2}\big)$ below the corresponding curve correspond to monotonous extrapolations. In particular, we observe that larger distances $\rhat$ allow for more smearing $\frac{8\tflow}{a^{2}}$ and for larger maximum lattice spacings. If we consider ensembles with lattice spacings $a\leq 0.06\,\fm$ and if we want to have monotonous extrapolations for distances $r\geq 0.14\,\fm$ ($\rhat\geq0.3$), we should choose $\frac{8\tflow}{a^{2}}\lessapprox 1$. In order to cover even smaller distances one needs smaller lattice spacings or less smearing.

\section{Comparison between gradient flow and stout smearing}

\begin{figure}
\centering
\includegraphics[width=0.495\textwidth]{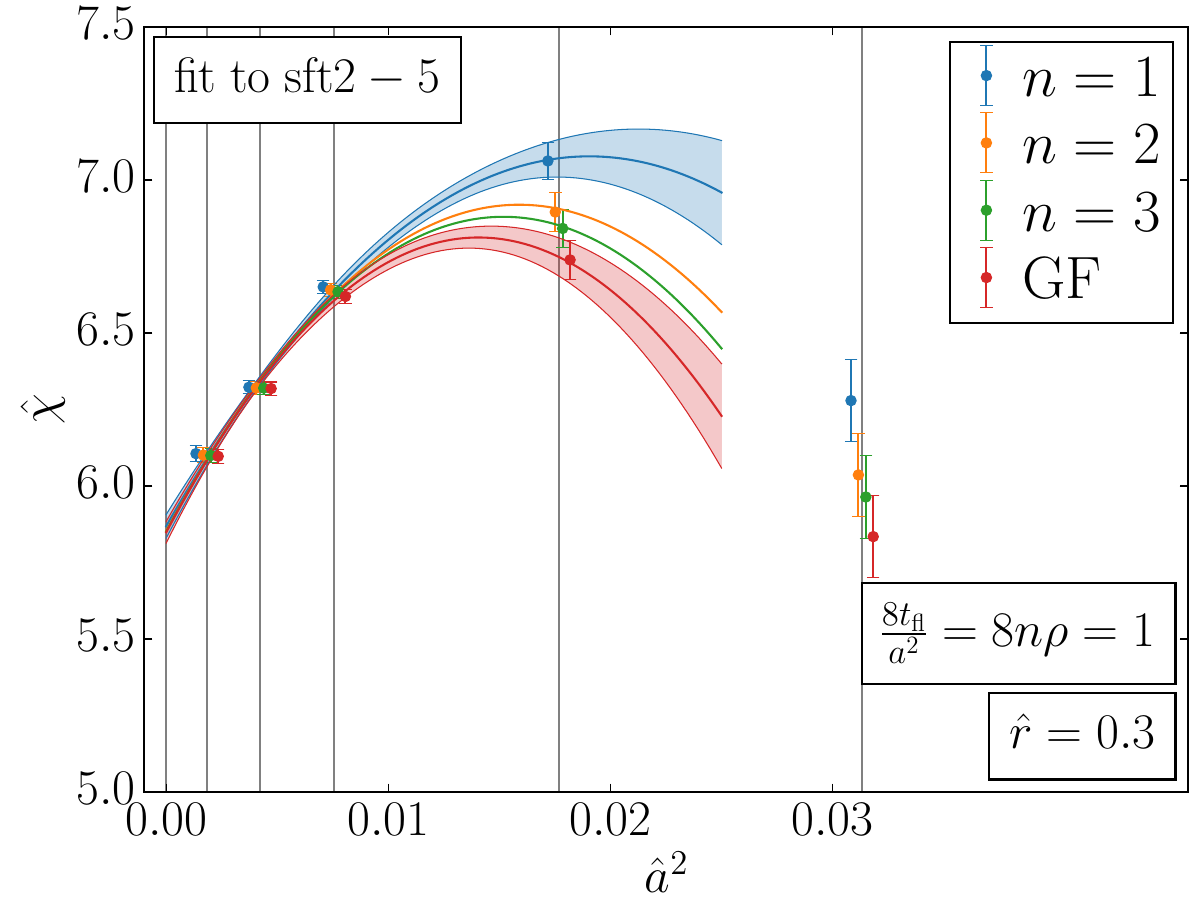}
\includegraphics[width=0.495\textwidth]{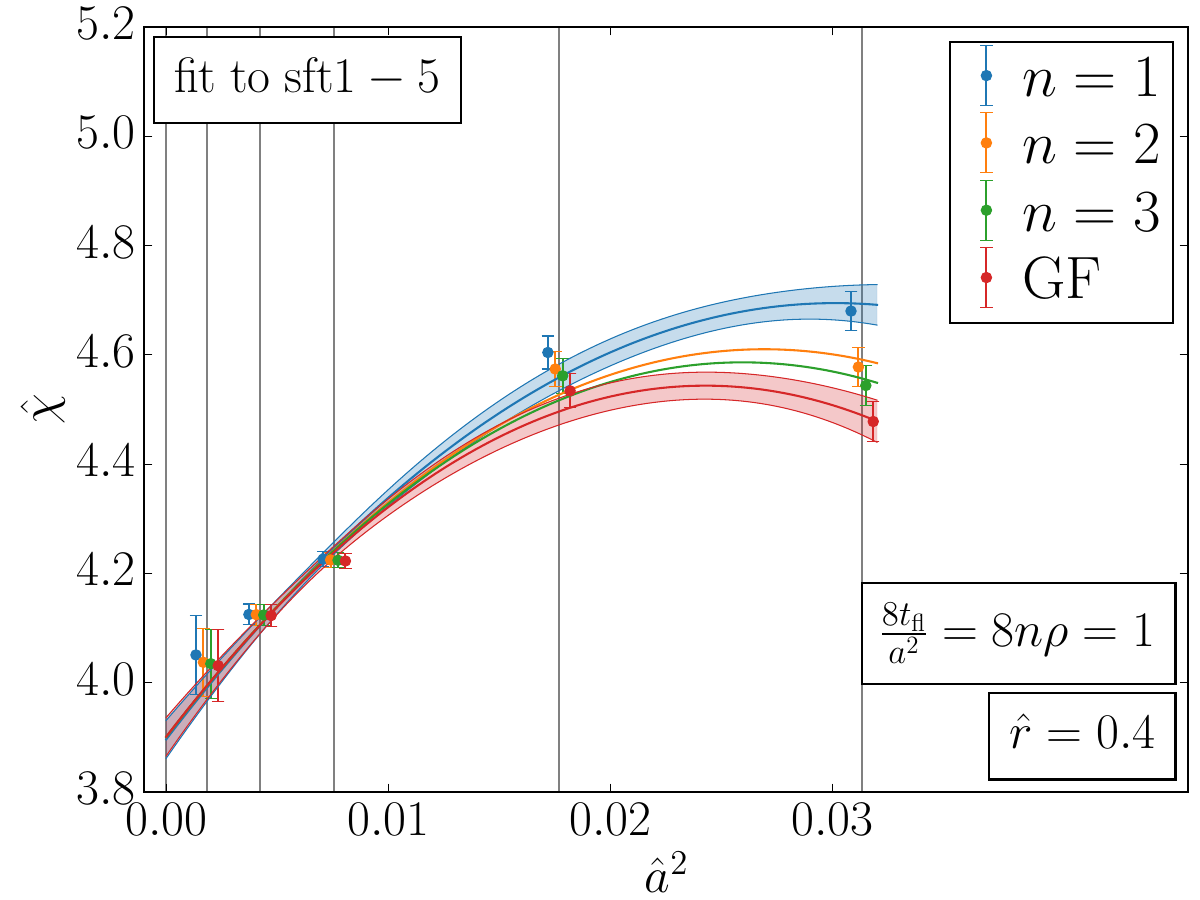}
\vspace{-0.6cm}
\caption{Continuum extrapolations of stout smeared Creutz ratio $\chihat$ at distances $\rhat=0.3/0.4$ ($r=0.14/0.18\,\fm$) as a function of the lattice spacing $\ahat^{2}$. Points have been shifted for better visibility.}
\label{fig:continuumstout}
\end{figure}
When simulating dynamical fermions, the application of the gradient flow on the gauge fields in the fermion Dirac operator is computationally demanding. One therefore employs stout smearing instead and only a small number of smearing iterations $n$ is applied. As in the limit of $n\to\infty$ stout smearing converges towards the gradient flow, both can be related to each other by $\frac{8 \tflow}{a^{2}}=8n\rho$, where $\rho$ is the stout smearing parameter. In \cref{fig:continuumstout} continuum extrapolations for stout and gradient flow smeared Creutz ratios are displayed with a smearing strength of $\frac{8 \tflow}{a^{2}}=8n\rho=1$. We observe that the absolute lattice artefacts grow when the gradient flow is approximated by stout smearing. However, when applying three stout iterations, the stout smeared Creutz ratio almost reproduces the gradient flow result within errors. With regard to the monotony criterion the location of the maximum of the stout smeared extrapolation is shifted to somewhat larger $\ahat^{2}$ compared to the gradient flow, i.e. our findings for gradient flow smearing form a conservative bound. In principle, even the application of one stout smearing iteration might be sufficient.

\section{Conclusion and Outlook}

We have studied the influence of gradient flow smearing on the continuum extrapolation of diagonal Creutz ratios $\chi(r,r)$ evaluated at various distances $0.14\,\fm \leq r \leq 0.28\,\fm$. We have found that the maximum tolerable smearing radius that still allows for a monotonous continuum extrapolation depends on the distance at which the Creutz ratio is evaluated. As expected, given a set of lattice gauge ensembles covering lattice spacings in a certain fixed range, for short distance observables less smearing is tolerable. The main result of our numerical investigation is summarised in \cref{fig:peaklocation}. Each curve yields the upper boundary of the region where a continuum extrapolation is monotonous, which is a minimum requirement for extrapolating data reliably to the continuum as we only have limited knowledge about the shape of the continuum extrapolation, in particular when higher order $a$ effects and logarithmic corrections~\cite{Husung:2017qjz} have to be considered. Therefore, a simple functional form is essential for a controlled extrapolation. For $\frac{8\tflow}{a^{2}}= 1$ we also studied the effect when the gradient flow is approximated by a small number $n$ of stout smearing~\cite{Morningstar:2003gk} iterations keeping $\frac{8 \tflow}{a^{2}}=8n\rho$ fixed. We found that the location of the maximum in the continuum extrapolation is shifted to somewhat larger $\ahat^{2}$ and that even a single stout smearing step with $\rho=\frac{1}{8}$ reproduces our qualitative findings. Although Creutz ratios give a measure of the force between static quarks and hence also have an implication for computations including fermions, it is important to corroborate our findings by considering various observables including fermions, where we fix the smearing to the found range.

\vspace{0.5em}
\begin{small}
We gratefully acknowledge the support of DESY where our computations were performed. AR would like to thank S. Schaefer and R. Sommer for a fruitful collaboration.
\end{small}

\bibliographystyle{JHEP}
\bibliography{proceedings.bib}

\end{document}